# Black Hole Entropy:

# a spacetime foam approach.

Fabio Scardigli[1]

Department of Physics, University of Milano, via Celoria, 16, 20133 Milano. Italy.

**Summary**: I introduce the spacetime foam structure by reviewing briefly the ideas of Wheeler, the topological fluctuations and the virtual black holes possibility. The contribution of Jacobson (the equation of state of the foam) is recalled.
In the second part, I introduce a model of spacetime foam at the surface of the event horizon of a black hole. I apply these ideas to the calculus of the number of states of a black hole, of its entropy and of other thermodynamical properties. A formula for the number of microholes on the surface of the event horizon is derived.
Subsequently, I extend the thermodynamical properties of the event horizon to thermodynamical properties of the space. Here I face with the problem of the maximum entropy contained in a space region of a given volume.
In the end, on the basis of the results previously obtained, I briefly treat the possibility of micro black holes creation by Unruh Effect.

gr-qc/9706030  11 Jun 1997

## 1.- Introduction.

During the last decade there have been many attemps to explain the Bekenstein-Hawking entropy [1] in terms of degrees of freedom attributed to a black hole (internal or external degrees of freedom, radiation degrees, etc.) [2],[3], [4],[5].
The approach developed in the present work is aimed at assigning the cause of the entropy of a black hole neither to inner fields nor to radiation fields near the event horizon, but to degrees of freedom of the spacetime region which forms the surface of the event horizon. More precisely, to the degrees of freedom of the space foam forming the geometrical surface of the event horizon.

---

[1] Private Address: Fabio Scardigli, via Europa, 20, 20097 S. Donato Milanese, Milano. Italy.
Fax: +39-2-5272698. E-mail: scardus@galactica.it

In the present model, the entropy does not depend on internal or external fields, but on the microstructure of the space.

In this view, the entropy of a black hole gives us vital informations on the foam microstructure of the space. The entropy is in this way connected with dynamical degrees of freedom of the spacetime itself. The degrees of freedom that have to be quantized are therefore the degrees of freedom of the space microstructure and not those of physical fields.

## 2.- Spacetime Foam.

The notion of spacetime foam was introduced by Wheeler in 1957 [6]: The vacuum fluctuations of the gravitational field increase in intensity at short wavelengths. If we extrapolate to the Planck region the standard results for the weak field, we find that the curvature fluctuations become so strong that are able to produce holes, tears and rips in the spacetime tissue, changing the topology by doing so. Wheeler imagines vacuum in a state of eternal agitation, with continual appearance of wormholes and other more complex structures of planckian width. Agitation is visible only at a planckian level. At a coarser level, the spacetime appears smooth. This set of topological microstructures continuosly in dynamical evolution was named by Wheeler *spacetime foam*.

The typical size of these space defects is the Planck length

$$l_p = \left(\frac{G\hbar}{c^3}\right)^{1/2} \tag{1}$$

At this scale, strong distorsions and rips appear in the spacetime topology, and virtual black holes may be formed with a typical gravitational radius of

$$R_g = l_p = 10^{-33} \, cm \tag{2}$$

Contributions to investigation and development of these ideas have been innumerable (see for example [7]).

Among the last papers in order of time there is that of T. Jacobson [8].

Jacobson' paper is a lively conceptual proof of the actual existence of some kind of topological complex spacetime structure at microscopic level.

In fact, the Einstein gravitational equation

$$G_{\mu\nu} = \chi T_{\mu\nu} \qquad (3)$$

is derived by assuming the proportionality of entropy and area of event horizon

$$dS = \eta \delta A \qquad (4)$$

together with the fundamental relation

$$\delta Q = TdS \qquad (5)$$

connecting heat Q (i.e. energy), entropy S, and temperature T.

It is demanded that the relation $\delta Q = TdS$ holds at local thermodynamic equilibrium. Besides, T is interpreted as the Unruh temperature on the event horizon

$$T = \frac{\hbar \kappa}{2\pi kc} \qquad (6)$$

Viewed in this way, the Einstein equation is an equation of state. It is obtained in the thermodynamic limit as a relation among thermodynamic variables and its validity depends on the existence of local equilibrium conditions. In fact the relation $\delta Q = TdS$ only applies to variations between nearby states of local thermodynamic equilibrium. This perspective suggests - in the words of Jacobson - that "it may be no more appropriate to quantize the Einstein equation than it would be to quantize the wave equation for sound in air".

The Einstein equation interpreted in this way appears as a macroscopic equation, which describes an average behaviour of an immense aggregate of microstructures, but it does not say anything about the single microstructure.

Exactly in the same way, using an appropriate set of macrovariables, the equation pV=NkT describes the average behaviour of a huge number of molecules whose individual behaviours are ignored. The possibility of deriving the Einstein equation as an equation of state, leads to believe that the system described by it, (pushing on the analogy with the gas), is composed by an enormous number of microsystems, which we can identify with the set of the topological distorsions and virtual microblack holes present at the planckian level, the so called spacetime foam. In other words the thermodynamical derivation of the Einstein equation gives a high degree of realism to a model in which the spacetime is viewed as a "foam" of planckian microstructures.

**3.- The Model.**

If a spacetime foam model presents such a good degree of likelihood, we can examine its predictive ability in regard to the entropy of black holes.

We must note, first of all, that the only region of the black hole that can be "seen" by a remote observer is the edge of the black hole, the horizon.

Therefore, *we assume that the entropy of the black hole must be attributed to the geometrical structure of the event horizon and to its possible micro degrees of freedom*.

The entropy of the hole has to be ascribed neither to inner fields (not observable), nor to the Hawking radiation trapped near the horizon. The entropy of a hole is actually definable in a way independent from the Hawking radiation.

On the grounds of the idea of spacetime foam, the spacelike surface $S^2$ "event horizon" is indeed thought, in the present model, as made up of various little topological structures (defects, rips, wormholes, microholes, tears, etc.) which I call for brevity "*topological cells*". The typical space dimension of these cells is, as said, the Planck length

$$l_p = \left(\frac{G\hbar}{c^3}\right)^{1/2} \tag{7}$$

The foam at the present stage is thought to be a two-dimensional layer covering the $S^2$ surface of the event horizon.

The shape of a single cell may be considered to be of a various kind. We can imagine that the cells have forms allowing a more or less regular covering of a spherical surface, for example a triangle, a square, a hexagon, an octagon, a circle.
The surfaces of these cells are respectively:

$$\frac{\sqrt{3}}{4}l_p^2, \quad 4l_p^2, \quad \frac{3}{2}\sqrt{3}\,l_p^2, \quad 2(1+\sqrt{2})l_p^2, \quad \pi l_p^2. \tag{8}$$

In general we consider cells of area $\alpha l_p^2$. The number of cells necessary to cover up all the horizon surface is

$$N = \frac{A}{\alpha l_p^2} \tag{9}$$

where $\alpha$ is a constant of the order of unity.

Suppose moreover that every cell may be occupied either by a space "smooth" or by a "rip" of the space tissue. In other words we can say that in every cell a micro black hole with a typical gravitational radius of $l_p$ may be present. As known, the energy necessary to create a "rip" (i.e. a microhole) is of the order of the Planck energy

$$\varepsilon_p = \frac{1}{2}\left(\frac{\hbar c^5}{G}\right)^{1/2} = 10^{19}\,GeV \tag{10}$$

The mass of a typical planckian microhole is the Planck mass

$$m_p = \frac{\varepsilon_p}{c^2} = \frac{1}{2}(\frac{\hbar c}{G})^{1/2} = 10^{-5} gr \qquad (11)$$

In this way, every topological cell has two states: "smooth" or "ripped". Such a physical model so described is basicly a two levels system. Each topological cell can stay either on the zero energy level - the "0" state in which the cell is smooth -, or on the level with energy $\varepsilon_p > 0$, the state "1" in which the cell is "ripped" ($\varepsilon_p$ = Planck energy).

To attribute to each topological cell a binary degree of freedom ("0"="smooth" or "1"="ripped") matches, for example, with the ideas expressed by 't Hooft in [9] :

"exp(S) is the total number of states available for a black hole with mass M (and energy $Mc^2$). This number is dominated by the black hole in such a way that there is roughly one Z(2) (binary) degree of freedom for each unit $l_p^2$ of surface area of the horizon". And again: "We have one physical Z(2) degree of freedom per unit $\delta\Sigma$ of surface area in the transverse direction".

This notion is resumed by Susskind in his work [10] and developed in the direction of the holographic hypothesis.

On the basis of the present model, the thermodynamical system considered is no more the black hole by itself, but the space foam building up the event horizon. We do not have to consider the internal degrees of freedom of a black hole as responsible for its entropy. The degrees of freedom that should be counted are indeed the number of states of the space foam in which consists the surface of the event horizon. We note also that in this model there isn't any physical membrane in place of the event horizon. The observable properties of the hole are not attributed to any physical membrane. This approach is different from the Membrane Paradigm. The entropy of the hole is actually the entropy of the space foam layer making up the event horizon.

**4.- Thermodynamics of the event horizon.**

In order to precise the thermodynamical theory of the horizon model outlined in the previous section, we have to assign the macrovariables which fix the macrostate of the thermodynamical system "event horizon". Successively we shall calculate the number of different ways in which a given microstate of the surface "event horizon" (compatible with the assigned macrostate of the system) can be obtained. The macro variables which characterize the space foam layer are:

N = overall number of topological cells = $\dfrac{A}{\alpha l_p^2}$ ;

U = total energy of the 2-level-thermodynamical-system of the topological cells.

Suppose to denote with $n_1$ and $n_2$ the numbers of cells which are respectively on the energy levels "0" and "1". The conditions on the numbers $n_i$ and on the energy U are

$$\sum_i n_i = N \qquad \sum_i n_i E_i = U \qquad (12)$$

where $E_1=0$, $E_2=\varepsilon_p$, $= \dfrac{A}{\alpha l_p^2}$ .

To calculate the number P of different ways in which a given microstate of the system "event horizon" (compatible with the assigned macrostate) can be constructed, we have to introduce another fundamental axiom. We claim that *each one of the two levels "0" and "1" has a double degeneracy*. In other words each one of the two levels with energy $E_1$, $E_2$ splits in two sublevels. It is straightforward [11] to see that the number of different ways belonging to the same microstate ($n_1$, $n_2$) is given by the formula

$$P = \dfrac{N!\, 2^N}{n_1!\, n_2!} \qquad (13)$$

I want to stress that the thermodynamical system here examinated is formed by a layer of space foam on the event horizon and it is characterized by the two macrovariables N and U, which fix the state of the system.

Observe, indeed, that a black hole macrostate is assigned when one gives the mass M, the charge Q, and the angular momentum J of the hole. In case of a Schwarzschild black hole N and M are linked by the relation

$$N = \frac{A}{\alpha l_p^2} = \frac{16\pi G^2 M^2}{\alpha l_p^2 c^4} = \frac{4\pi}{\alpha}\left(\frac{M}{m_p}\right)^2 \qquad (14)$$

The total energy U of the space foam layer can be assigned with a certain degree of arbitrariety. The absolute limits on the energy of the slice of space foam are, obviously:

$$0 \leq U \leq N\varepsilon_p = \frac{4\pi}{\alpha}\left(\frac{M}{m_p}\right) M c^2 \qquad (15)$$

From the fact that, usually, $\frac{M}{m_p} \gg 1$, we can derive that a "realistic" limit on the energy U is

$$0 \leq U \leq Mc^2 \qquad (16)$$

where M is the mass of the hole. The energy of the slice of foam cannot exceed, evidently, the total mass of the hole itself.

The most probable distribution can be obtained in the usual way (see e.g. Schroedinger [11]). We look for the maximum of the expression

$$\log P = N \log N + N \log 2 - \sum_i n_i \log n_i \qquad (17)$$

by varing the numbers $n_i$ under the conditions (12).

Applying the method of lagrangian multipliers we get the free maximum of the expression

$$\log P - \alpha \sum_i n_i - \beta \sum_i n_i E_i \qquad (18)$$

with arbitrary α and β.

The $n_i$ which maximize P are

$$n_i = \exp(-1-\alpha-\beta E_i) \qquad (19)$$

If we define the partition function of the system as

$$Z = \sum_i \exp(-\beta E_i) \qquad (20)$$

we arrive finally to the most probable distribution

$$n_i = \frac{N}{Z}\exp(-\beta E_i) \qquad (21)$$

The total energy U of the system "event horizon" results

$$U = \sum_i n_i E_i = \frac{N}{Z}\frac{dZ}{d\beta} \qquad (22)$$

The function $Z(\beta)$ can be easily computed in our two level system (event horizon)

$$Z = \sum_i \exp(-\beta E_i) = 1 + e^{-\beta \varepsilon_p} \qquad (23)$$

Of course $\beta$ is linked with the absolute temperature of the system as

$$\beta = \frac{1}{kT} \qquad (24)$$

Because the horizon is in thermal equilibrium with the Hawking radiation [12], it should be

Reminding

$$T = T_{Hawking} = \frac{\hbar c^3}{8\pi k G M} \qquad (25)$$

we can compute

$$\varepsilon_p = \frac{1}{2}(\frac{\hbar c^5}{G})^{1/2} = 10^{19} GeV \qquad (26)$$

$$\beta_H \varepsilon_p = \frac{\varepsilon_p}{k T_H} = 2\pi \frac{M}{m_p} \qquad (27)$$

where M is the black hole mass and $m_p$ is the Planck mass. In our two level system we can write for the occupation numbers $n_i$

$$n_1 = \frac{N}{Z} = N\frac{1}{1+e^{-\beta_H \varepsilon_p}}$$

$$n_2 = \frac{N}{Z}e^{-\beta_H \varepsilon_p} = N\frac{1}{1+e^{\beta_H \varepsilon_p}}$$

(28)

For a solar mass black hole we have

$$\beta_H \varepsilon_p = 2\pi\frac{M}{m_p} = 2\pi\frac{10^{33}}{10^{-5}} = 10^{38} >> 1$$

$$N = \frac{4\pi}{\alpha}(\frac{M}{m_p})^2 = 10^{77}$$

(29)

So in this case

$$Z = 1+\exp(-\beta_H \varepsilon_p) = 1 \qquad (30)$$

with excellent approximation;
and, as one easily sees, we have

$$n_1 = N; \qquad n_2 = 0 \qquad (31)$$

that is to say, there are no "rips" or "microholes" on the surface of the event horizon of a solar mass black hole at thermodynamic equilibrium.

Observe that the expression (28) of $n_2$ allows us to give an estimation of the number of microholes created on the event horizon as a result of the agitation of space foam.
We can ask which is the critical value of the mass M of the hole that maximizes the number of "rips" (i.e. microholes) on the surface of the event horizon. Finding the maximum of the function $n_2(M)$, we get $M \approx 1/3\ m_p$. This value does not have actually any physical meaning because it does not make sense to allow the presence of microholes with a mass lower than the Planck mass. They should have a gravitational radius shorter than the Planck length and they would get mixed up with the basic topological structure of the spacetime, the quantum foam, in this way becoming indistinguishable from it.

The expression of the total energy of the system "event horizon" is

$$U = -\frac{N}{Z}\frac{dZ}{d\beta} = N\varepsilon_p \frac{1}{1+e^{\beta_H \varepsilon_p}}. \qquad (32)$$

We see again here that in case of a solar mass black hole U=0. The total energy of the system "event horizon" is therefore zero with a very good approximation.
The average energy ascribables to every topological cell is

$$E_{ave} = \frac{U}{N} = \varepsilon_p \frac{1}{1+e^{\beta_H \varepsilon_p}}. \qquad (33)$$

We want to evaluate finally the expression of the entropy S of the event horizon (and therefore of the hole).
Following Boltzmann, we have

$$S = k\log P = k(N\log N + N\log 2 - \sum_i n_i \log n_i) =$$
$$= kN\log 2 + k\beta_H U + kN\log Z \qquad (34)$$

that is

$$S = kN\log 2 + \frac{U}{T_H} + kN\log Z. \qquad (35)$$

Reminding that in the present case it is

$$N = \frac{A}{\alpha l_p^2}; \quad U = N \varepsilon_p \frac{1}{1 + e^{\beta_H \varepsilon_p}}; \quad Z = 1 + e^{-\beta_H \varepsilon_p} \quad (36)$$

we can write

$$S = kN \log 2 + kN \beta_H \varepsilon_p \frac{1}{1 + e^{\beta_H \varepsilon_p}} + kN \log(1 + e^{-\beta_H \varepsilon_p}) . \quad (37)$$

We recover in the first term the "static" or "geometric" contribute to the entropy of the horizon (and so of the hole). The other two terms depend on the temperature of the hole and on $\varepsilon_p$.

For a black hole of one solar mass immediately results that the leading term in entropy is, as expected,

$$S = k \frac{A}{\alpha l_p^2} \log 2 \quad (38)$$

That is, the prevailing contribute is due to the "geometric" (or "static") term of the entropy.
The other two terms become important only for planckian size black holes.

**5.- Extension of the thermodynamical properties of the event horizon to thermodynamical properties of the spacetime.**

In this section we shall try to extend the thermodynamical properties previously derived for the surface of the event horizon to the whole physical space. Of course, also here we shall treat with a highly speculative model, but we are driven to do this because the spacetime foam is (or should be) a concept which applies to three (or even four) dimensional space.
A flat space region $A \subset R^3$ of volume V is subdivided in little topological cells by a 3-dimensional cubic lattice, each cell of which has a volume of the order of $l_p^3$. Each cell (as on event horizon surface) may be "smooth" or "ripped" (that is,
a micro black hole may, or may not, be present). The state "0" has zero energy; the state "1" (ripped) has an energy $\varepsilon_p$ necessary to create a microhole. This model of space foam has a direct filiation with the originary model of spacetime foam of Wheeler.
Here the foam is thought to be a 3-dimensional lattice (not necessarily regular) pervading the whole physical space.
This model agrees with the results of Hawking's work on "Spacetime Foam" [7]. In that work, Hawking shows (with an esteem of the path integral in the one-loop approximation) that the dominant contribution to the number of states with a given volume V comes from the metrics with Euler numbers $\chi$ of the order V, i.e. metrics that represent one gravitational instanton (i.e. one microhole) per unit Planck volume.
The number of topological cells contained in the volume V is

$$N = \frac{V}{l_p^3} \qquad (39)$$

We suppose again that each cell is characterized by two states with energies respectively $E_1=0$ and $E_2=\varepsilon_p$, and each one of the two levels has two sublevels (double degeneracy).
The macrovariables which fix the macrostate are, of course, the number N and the energy U of the system of topological cells. As in section 4 we can evaluate the number of different ways in which the microstate $(n_1, n_2)$ can be achieved:

$$P = \frac{N! \, 2^N}{n_1! \, n_2!} \qquad (40)$$

The most probable distribution is computed in the usual way, working out the maximum of the expression

$$logP \tag{41}$$

when numbers $n_i$ vary under the conditions

$$\sum_i n_i = N; \qquad \sum_i n_i E_i = U. \tag{42}$$

We find in this way the occupation numbers of the two levels 0 and 1

$$n_i = \frac{N}{Z}\exp(-\beta E_i) \tag{43}$$

with $E_1=0$ energy of the state "0"; $E_2=\varepsilon_p$ energy of the state "1". As before, the partition function Z is defined as

$$Z = \sum_i e^{-\beta E_i} = 1 + \exp(-\beta_U \varepsilon_p) \tag{44}$$

We can call $n_i$ the "occupation numbers of the thermodynamical system of the space foam". $\beta$ is identified with the reciprocal of the absolute temperature of the system of the "topological cells of space foam". The appropriate temperature cannot be choosen again as the Hawking temperature, because here we are not near a black hole event horizon. Nevertheless, the space region of volume V has a temperature of thermodynamic equilibrium that can be coherently identified with the Unruh temperature of the vacuum in the reference frame describing V. This because we can think that the system "topological cells of space foam" is in thermodynamic equilibrium with the zero point radiation, whose temperature in an accelerated system is the Unruh temperature [13]. Therefore we put

$$\beta_U = \frac{1}{k\,T_U} \quad with \quad T_U = \frac{\hbar a}{2\pi k c} \tag{45}$$

where a is the acceleration of the system.

We can evaluate the factor $\beta_U \varepsilon_p$ which appears for example in $n_2$ and in Z=1+exp($-\beta_U\varepsilon_p$). Reminding the expression for $\varepsilon_p$ we easily find

$$\beta_U \varepsilon_p = \frac{\varepsilon_p}{k T_U} = \frac{\pi c^2}{a} \frac{1}{l_p} \qquad (46)$$

The total energy of the system of the space foam contained in a volume V is

$$U = -\frac{N}{Z}\frac{dZ}{d\beta} = N \varepsilon_p \frac{1}{1+e^{\beta_u \varepsilon_p}} \qquad (47)$$

but now we have

$$N = \frac{V}{l_p^3} \qquad (48)$$

Eventually, we observe that the model of space time now proposed implies that a volume V of space foam has an entropy equal to

$$S = k \log P = kN \log 2 + k \beta_U N E_{ave} + kN \log Z \qquad (49)$$

where

Observe that $\beta_U \varepsilon_p$ is enormous for accelerations not too large. Even if a is of the order of the acceleration experimented by an electron orbiting a proton in an hydrogen atom we derive that $\beta_U \varepsilon_p = a l_p^{-1} \times \frac{e^2}{m_e r^2} 10^{30} \approx 2.53 \times 10^{24}$. From which follows $\qquad (51)$

$$\beta_U E_{ave} = 0; \quad \exp(-\beta_U \varepsilon_p) = 0; \quad Z = 1. \qquad (52)$$

Therefore for "every day" accelerations we have:

$$S = kN(\log 2 + \beta_U E_{ave} + \log Z) = kN \log 2 \qquad (53)$$

The leading term, that is the most important contribution to entropy, is again the geometric (or static) term. And that term is proportional to the volume:

$$S = kN \log 2 = k \frac{V}{l_p^3} \log 2 \qquad (54)$$

Therefore the maximum entropy of a region of space or, that is the same, the maximum entropy of the space foam contained in a volume V, is *proportional to the volume* of the region considered. This entropy is clearly at maximum, since it is not possible to imagine a microstructure smaller than the element of a cubic lattice with side $l_p$. More little microholes would be mixed up, dissolving in the spacetime foam.

Note that the entropy results proportional to the volume and not to the area of the considered space region. This fact disagrees with the ideas exposed by Bekenstein [14],[15] who claims that the entropy goes as the area and not as the volume of the considered region.
Susskind, in his article " The world as a hologram" [10], asserts that if the energy density is bounded, the maximum possible entropy of a space region is proportional to the volume of the region. "It is hard - Susskind says - to avoid this conclusion in any theory in which the laws of nature are reasonably local". Nevertheless, Susskind stresses that there are good reasons to believe that the correct result in the quantum theory of gravity is that the maximum entropy is proportional to the area and not to the volume of the region. Susskind reports a Bekenstein's argument to support this thesis: Suppose that the region of volume V has an entropy S÷V. By compressing matter in that region, it is possible to give rise to a black hole that occupies all the region. But the entropy of the hole is S∝A (A area of the region). So, by throwing in (additional) matter in the region V we should be able to create a hole with S ∝ A < V, therefore violating the Second Law. Bekenstein gives then other direct reasons [2] which all lead to a superior limit of

$$S \leq k \frac{2\pi RE}{\hbar c} \qquad (55)$$

For a Schwartzshild black hole E∝M∝R, therefore S∝R$^2$, that is S is proportional to the area and not to the volume of the region. Without entering the demonstrative details, I want to say that the argument of Bekenstein, quoted before, makes reference to the superior limit of the entropy of space regions *trapped by an event horizon*. I suspect that this limit, certainly true, is nevertheless applicable only to space regions internal to an event horizon, that is, to black holes. The superior limit for the entropy of "free" space regions (i.e. not trapped by a horizon) would seem more reasonably fixed by the proposed model. The model assumes a cubic lattice microstructure of the spacetime and leads in a natural way to an entropy proportional to the volume V.

From this model of 3-dimensional foam we may derive the proportionality between entropy and area in the case of a black hole (i.e. in presence of horizons). We observe that in regions without any event horizon the model foresees S ∝ V everywhere. Since only difference measures may be taken for entropy values (but not absolute measures), the entropy of ordinary flat space results zero. But, if a space region is trapped behind an event horizon, it results not observable. Therefore it is possible to attribute to such a region a single internal state and, as a consequence, a null entropy (black hole internal entropy). The entropy of the space foam observable out of the hole is, on the contrary, different from zero. The external observer "sees" a surface entropy that is the visible residue of the entropy of the space foam surrounding the hole. More precisely, the only observable region is the horizon. It is on the horizon that the lattice properties of the space display themselves as a system composed of $N=A/l_p^2$ microcells, able to assume $2^N$ different states and therefore characterized by an entropy of $S \propto N = A/l_p^2$. Introducing the notion of *stretched lattice*, we can better understand the reason for which the contribute to the entropy of the black hole comes from the topological cells put in contact with the horizon, while those far from the horizon do not give any contribute.

---

[2] Particularly the calculus of the maximum entropy that can be owned by a fixed quantity of radiation closed in a box.

From the Schwartzschild metric

$$ds^2 = (1-\frac{R_S}{r})c^2 dt^2 - (1-\frac{R_S}{r})^{-1} dr^2 - r^2 d\Omega^2 \qquad (56)$$

we see that the length of the radial element is infinitely "stretched" near the horizon:

$$(1-\frac{R_S}{r})^{-1} dr^2 \to \infty \quad for \quad r \to R_S \qquad (57)$$

This involves a lensing effect for the far observer: the topological cells in contact with horizon surface are "magnified" and give the prevailing contribute to the entropy of the hole in respect to the other cells of 3-dimensional cubic lattice. Evidently the number of cells in contact with the horizon is $N = A/l_p^2$. Therefore we can apply all the preceding arguments to derive an entropy of the black hole proportional to the area.

**6.- Microholes Creation by the Unruh Effect.**

The occupation number

$$n_2 = \frac{V}{l_p^3} \frac{1}{Z} e^{-\beta_U \varepsilon_p} \qquad (58)$$

is the number of topological cells that are on the level "1" and each "1"-cell corresponds to a planckian microblack hole. As a consequence the quantity

$$\frac{n_2}{V} = \frac{1}{l_p^3} \times \frac{1}{Z} e^{-\beta_U \varepsilon_p} \qquad (59)$$

may be interpreted as the number of microblack holes per unit volume, created in a region of volume V, because of the fact that the region is accelerated and the quantum foam is excited to Unruh temperature. We know that $\beta_U$ is linked to Unruh temperature by the relation

$$\beta_U = \frac{1}{k T_U} = \frac{2\pi c}{\hbar a} \qquad (60)$$

and the factor $\beta_U \varepsilon_p$ is

$$\beta_U \varepsilon_p = \frac{\pi c^2}{a} \frac{1}{l_p} \qquad (61)$$

where $a$ is the acceleration undergone by the reference system containing the region of volume V.
We have seen that, for not too large accelerations, the factor $\beta_U \varepsilon_p$ is enormous. If $a$ is the centripetal acceleration of the electron in the hydrogen atom ($a = 2.5 \times 10^{24}$ cm/sec$^2$) we have $\beta_U \varepsilon_p = 10^{30}$. Therefore, in the rest frame of the electron the number of planckian microholes created by the thermal excitation of the quantum foam due to the Unruh effect is practically zero:

$$\frac{n_2}{V} = \frac{1}{l_p^3}\frac{1}{Z}e^{-\beta_U \varepsilon_p} = 10^{99}\exp(-10^{30}) = 0 \qquad (62)$$

For ultrarelativistic electrons (or others particles) confined in a storage ring, the acceleration in the rest frame is given by [16][3]

$$a = \frac{\gamma^2 c^2}{R} \qquad (63)$$

and therefore the factor $\beta_U \varepsilon_p$ becomes

$$\beta_U \varepsilon_p = \pi \frac{R}{\gamma^2 l_p} \qquad (64)$$

where R is the radius of the ring. Particularly, the factor $\beta_U \varepsilon_p$ can be made small by taking γ large and R short.
In case of 100 GeV electrons at LEP we have

$$\gamma = \frac{E}{m_e c^2} = \frac{100\,GeV}{5\times 10^{-4}\,GeV} = 2\times 10^5 \qquad (65)$$

and R = 3.1×10⁵ cm. Therefore

$$\beta_U \varepsilon_p = 1.5\times 10^{28} \qquad (66)$$

---

[3] In fact centripetal acceleration is $a = \frac{v^2}{R}$. Since a rule moving along a circumference of radius R contracts by a factor γ in the moving direction (l' = l/γ) we have for the actual velocity v' = l'/T = l/γT that is v=v'γ. But v'=c for ultra relativistic electrons. Therefore a = γ²c²/R. Actually, a rigorous treatment gives the correct relation a = (γ² -1)c²/R. But for ultra relativistic electrons γ>>1.

The number of microholes per unit volume created for Unruh excitation of the quantum vacuum is, in this case (Z=1),

$$\frac{n_2}{V} = \frac{1}{l_p^3} e^{-\beta_u \varepsilon_p} = 10^{98} \exp(-10^{28}) = 0 \tag{67}$$

Thus, we do not have a great production of microholes in a ring like LEP !!!

Let us look for the characteristics that a storage ring (or a collider) should have to take electrons to energies great enough to excite the vacuum quantum foam, by Unruh effect, up to Planck energy, allowing in this way the production of micro black holes.

If we want to have at least 1 microhole for every cm$^3$ (with $R_{SCH}=10^{-33}$ cm, mean life = 2000 Planck Time) we have to set

$$\frac{1}{l_p^3} e^{-\beta_u \varepsilon_p} = 1 \tag{68}$$

from which

$$\frac{R}{\gamma^2} = -\frac{3}{\pi} l_p \log l_p = 1.15 \times 10^{-31} \, cm \tag{69}$$

Supposing to maintain a radius of the order of LEP radius (R=3×10$^5$ cm) we obtain for γ

$$\gamma = 1.6 \times 10^{18} \tag{70}$$

To accelerate electrons at this γ, we must reach energies of the order of

$$E = \gamma m_e c^2 = 1.6 \times 10^{18} \times 5 \times 10^{-4} = 8 \times 10^{14} \, Gev \tag{71}$$

As one may see, this is an energy lower than the fatidical Planck threshold ($10^{19}$ GeV), yet it is again very much distant from the energies reached with up-to-now accelerators ($10^{3}$ GeV) or with the next generation machines ($10^{4}$ GeV).

It does not seem possible to obtain microholes in our accelerators (at least by Unruh effect) in an immediate future, unless unexpected and big improvements of the accelerator technology.

**7.- Acknowledgments.**

I wish to thank L. Girardello for the useful discussion on the arguments of this paper.

======== OOO ======== OOO ========

**REFERENCES**

[1] J.D. Bekenstein, Lett. Nuovo Cimento, 4, (1972), 737.
    Bardeen, Carter, Hawking, Commun. Math. Phys. 31, (1973) 161.

[2] W.H. Zurek, K.S. Thorne, Phys.Rev. Lett. 54, (1985) 2171.

[3] G. 't Hooft, Nuclear Phys. B 256 (1985) 727-745.

[4] V.Frolov, I.Novikov, Phys.Rev. D 48 (1993) 1607.
    V.Frolov, I.Novikov, Phys.Rev. D 48 (1993) 4545.

[5] J.W. York, Phys.Rev. D, 28, (1983) 2929.

[6] J.A. Wheeler, Geometrodynamics, (Academic Press, New York, N.Y.) 1962.

[7] S.W. Hawking, Nuclear Phys. B 144 (1978) 349-362.
    T. Regge, Nuovo Cimento, vol. 19, (1961) 558-571.

[8] T. Jacobson, " Thermodynamics of Spacetime: the Einstein Equation of State", gr-qc/9504004; UMDGR-95-114, (1995).

[9] G. 't Hooft, Physica Scripta, vol. T36, (1991) 247.

[10] L. Susskind, J.Math.Phys., 36,(1995) 6377.

[11] E. Schroedinger, Statistical thermodynamics (Cambridge U.P., 1952).
M. Alonso, E. Finn, Statistical Physics, Addison Wesley.

[12] S.W. Hawking, Comm. Math. Phys. 43, (1975) 199.

[13] W.G. Unruh, Phys.Rev. D, 14, (1976) 870.

[14] J.D. Bekenstein, Phys.Rev. D, 23, (1981) 287.

[15] J.D. Bekenstein, Phys.Rev. D, 49, (1994) 1912.

[16] J.S. Bell, J.M. Leinaas, Nuclear Phys. B 212 (1983) 131.